# Improving the CSIEC Project and Adapting It to the English Teaching and Learning in China

Jiyou Jia[1], Shufen Hou[2], Weichao Chen[3]
*1,3 Department of Educational Technology, Peking University, China*
*jyjia, st510@gse.pku.edu.cn*
*2 Department of English, Henan Agriculture University, China*
*houshufen@csiec.com*

## Abstract

*In this paper after short review of the CSIEC project initialized by us in 2003 we present the continuing development and improvement of the CSIEC project in details, including the design of five new Microsoft agent characters representing different virtual chatting partners and the limitation of simulated dialogs in specific practical scenarios like graduate job application interview, then briefly analyze the actual conditions and features of its application field: web-based English education in China. Finally we introduce our efforts to adapt this system to the requirements of English teaching and learning in China and point out the work next to do.*

## 1. Introduction

We have designed and implemented the framework of a web-based human-computer-communication system with natural language for foreign language learning, CSIEC (Computer Simulator in Educational Communication) [1]. It consists of the NLML (Natural Language Markup Language) which structurally labels the grammar elements, NLOMJ (Natural Language Object Model in Java) which represents the grammatical elements in Java, NLDB (Natural Language Database) which stores the historical discourse, and the CR (Communicational Response) mechanism considering the discourse context, the world model and the personality of the users and of the system itself comprehensively. The structure of this system is shown in Figure 1.

Since its birth three years ago this system has been applied in a website and browsed freely by the English learners in China at any time. Without amounts of advertisements the website homepage has been clicked more than 40 thousand times. On the other hand this project has arrested the researchers' attention from the fields such as educational technology, and CALL (Computer Assisted Language Learning), as well as the reports from mass media [2]. Meanwhile we have been continuing working not only on realizing the system from the original prototype to a practical interactive web-based E-Learning system and improving it, but also on the research of its application on the special context of English teaching and learning in China.

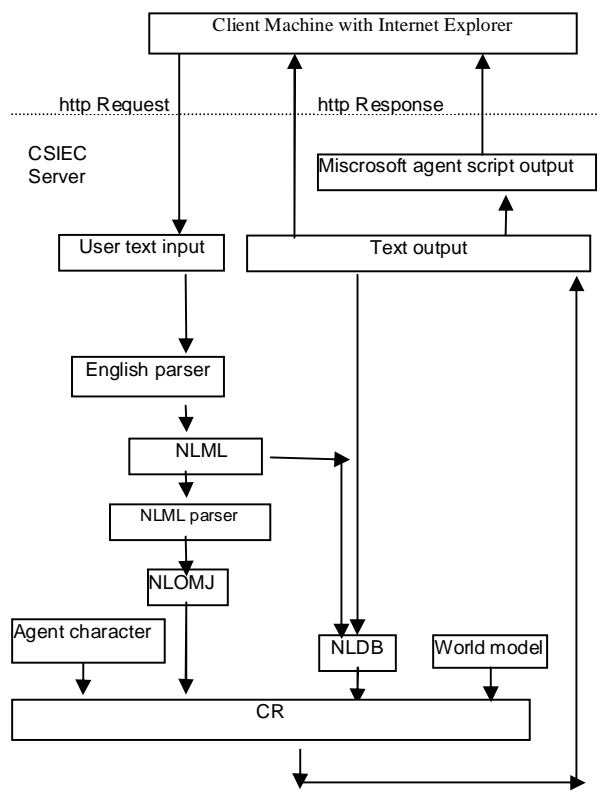

*Figure 1* The structure of CSIEC system

## 2. The Design of New Microsoft Agent Characters Representing Different Virtual Chatting Partners

With the function of communicational response the CSIEC system can be used to work as a virtual chatting partner in the Internet for the English learners. This is its major application field for English learning. In the current version there is still only one user input method, i.e. text input, because the speech recognition technology state of the art is not so satisfactory yet. Apart from the pure text output we use also Microsoft agent character technology [3] to produce synthesized speech output, because its voice is lifelike, the figures, movements and actions can be designed very vivid and attractive, and it can also dynamically display the spoken text, what is conducive to the aural understanding by language learning.

As every user, a real human being, has his (or her) character and personality, he (or she) may choose different type of chatting pattern by online-chatting. For example some users prefer one partner who can quietly listen to them; on the contrary some others hope the chatting partner can tell them more stories, jokes or news.

For the sake of dialog personalization we have designed five Microsoft agent characters using the Microsoft Agent Character Editor. Their names, the main pictures of the animation associated with the showing state (every character must have a showing state in order to be able to appear on the screen), the chatting patterns, some dialog examples between a user and the selected agent character, and simple explanations based on agent theory are shown in the following paragraphs one by one.

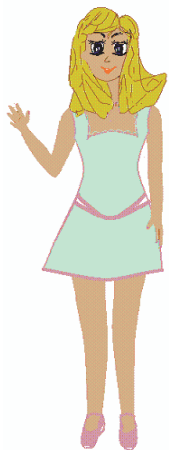

This sexy girl (agent character) is Christine. She always tells the user stories, jokes and world news. The following is an example dialog between Christine and a user named John.

John: Hello, Christine.

Christine: Good morning, John. I have known your name from our previous dialog. This story is a joke.

John: Please then.

Christine: Little Johnny came running into the house and asked, 'Mommy, can little girls have babies?'

John: what is the answer?

Christine: 'No,' said his mom, 'Of course not.'

John: ha, ha!

Christine: Little Johnny then ran back outside and his mom heard him yell to his friends, 'It's okay, we can play that game again!'

John: Funny!

Christine: That is all I know about the joke of Baby Talk.

Etc.

The example dialog shows that this kind of chatting agent represented by Christine is an active agent.

The next handsome boy (agent character) is Stephan. He always quietly listens to the user pouring out his (or her) heart, and the users like to cry on his shoulder. The following is an example dialog between a user named Yang and Stephan.

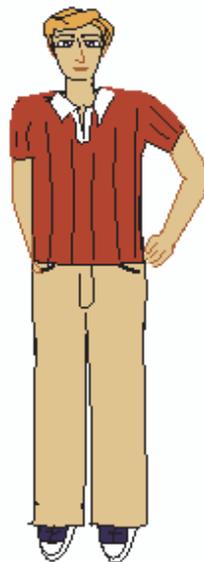

Stephan: Good morning, Yang. Happy to meet you again!

Yang: I want to cry these days.

Stephan: oh? Why?

Yang: I have failed in my final examination.

Stephan: So terrible?

Yang: Yes, this course is extremely important one in my major.

Stephan: then?

Yang: I must learn it again in next semester.

Stephan: oh.

Etc.

The example dialog shows that this kind of chatting partner represented by Stephan is a reactive agent. It says little about itself or other things.

The slim girl (agent character) is Emina, and always asks the user all kinds of curious questions related with the user input text. The following is an example dialog between a user named Andreas and Emina.

Andreas: Hi, Emina.

Emina: Hello, Andreas! Do you like the Internet?

Andreas: Yes.

Emina: Oh, you like the Internet. Why do you like the Internet?

Andreas: because I can get any information I need.

Emina: ha, you like Internet because you can get any information you need.

Andreas: yes. I am very happy this week.

Emina: Were you happy before?

Etc.

The example dialog shows that this kind of chatting partner represented by Emina is a proactive agent. It likes to obtain another's information by asking questions.

The next strong boy (agent character) is Christoph. He makes comments, suggestions and advices on the user's input text. The following is an example dialog between Christoph and a user named Anna.

Anna: Hi.

Christoph: Happy weekend, Anna. Do you watch TV?

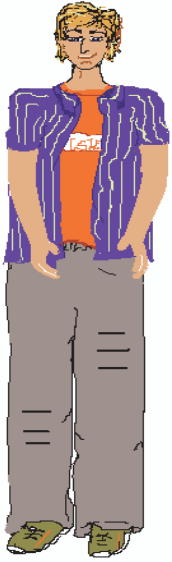

Anna: Yes.

Christoph: Oh. You watch TV. But watching TV wastes the students too much time, and even hurts their eyes.

Anna: I watch TV to learn English.

Christoph: The best way to learn English is to use it as often as possible, fox example, to chat with me everyday.

Anna: thank you. I will face an English examination tomorrow.

Christoph: Everything ready? Believe yourself.

Etc.

The example dialog shows that this kind of chatting partner represented by Christoph is also a proactive agent.

Besides the four predefined chatting partners represented by the Microsoft agent characters there is still one comprehensive virtual chatting partner, which gives the user responses corresponding to both the input text and the discourse context. It can be accessed either by the following girl (agent character) Ingrid or through the pure text chatting version. An example dialog between a user named Carl and this chatting pattern is shown in the following paragraph.

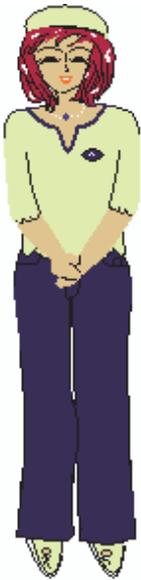

Carl: can you sing a song.

Ingrid: yes. I can sing a song. You have said you can't sing a song according to our previous dialog.

Carl: It is right. Please sing a song for me.

Ingrid: Daisy, Daisy.

Give me your answer do.

I am half crazy.

… (The other words of the song are omitted)

Carl: you are very clever.

Ingrid: Yes, I am smart because I have a lot of knowledge in my brain. Are you clever?

Carl: No.

Ingrid: Oh, You are not clever. To be clever one must learn and learn during all the life.

Carl: yes. I want to tell you a story now.

Ingrid: please.

Etc.

The example dialog shows that this kind of chatting partner represented by Ingrid is a hybrid agent.

Through the above description we can see that the users can select different kinds of virtual chatting partners.

It should be noticed that all these various characters make same responses to the interrogative and imperative sentences from the users. On other words they should try to answer user questions, and meet with a response to the users' demand, order, request or ask. The differences among these characters lie on one side mainly on their response to the statement sentences. For example Emina makes questions about a statement sentence by transferring the sentence tense, or adding question auxiliary adverb to the whole sentence, but the Christoph gives comments, suggestions or advice based on the verb and other main parts in the statement sentences, no matter what tense or forms they are in. On another side the inherent chatting redlines of the agents are diverse.

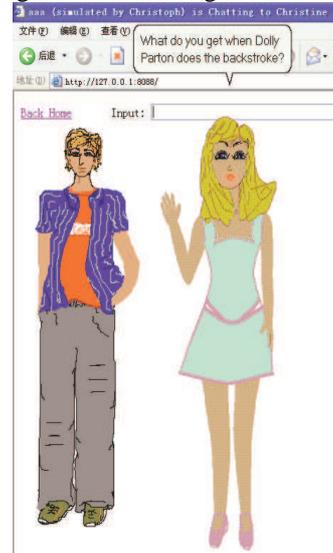

*Figure 2* A snapshot of a client using Internet Explorer to chat with the virtual partner Christine

Furthermore the users can also choose one from the five Microsoft agent characters (Christine, Stephan, Ingrid, Emina, Christoph) to represent themselves on behalf of their preference (sex, figure, clothes, etc.), while the five ones appear obviously differently, as shown above. By simulating one user the agent character has not the fixed chatting pattern described above any more, however, it speaks the text inputted by the user, and therefore looks like a stand-in of the user on the computer screen during the human-computer interaction with natural language. In Figure 2 there is a snapshot of the dialog between Christoph representing a user and Christine representing the virtual chatting partner.

## 3. Limited dialog simulation in special scenarios such as job application interview

Up to now we have discussed the chatting mechanism in the unlimited dialog between the user and the agents. "Unlimited" means here that neither the theme, topic, time, nor the duration is limited. In other words the users can chat with the agent about anything. The unlimited chatting benefits the user whose English is fluent or who at least is good at written English, and whose character is extroversive or conversational. But the user whose English (at least the written English) is poor, or who is introversive, has only little to talk with the virtual chatting partner. So the dialogs between the agent and the latter users are mostly very short according to our measure method in [4]. For these users an instructive dialog in a specific scenario guided by the agent is more useful.

In actual English instruction these two kinds of dialogs take place often in different learning stages. The limited dialog guided by the teacher is appropriate for the junior learners, school students, and happens often in the classroom. The unlimited dialog fits the advanced learners, college and university students, and take place usually in the informal occasions like English corners, etc.

Based on this consideration we have designed some limited dialog simulations in special scenarios such as graduate job application interview, university entrance interview, and so on. Here we explain the example of graduate job application interview in details, as it is very pragmatic for the university and college students, job searchers, etc., who are the main users of the current system and wish to win a position in an international enterprise in which the English skills are prerequisite.

In this case the topics between the interviewer and interviewee can be normally classified into three categories: college and specialty, student lives, and reasons for this job application. The dialogs about these topics can also be organized in this order. In the official interview the interviewer asks the interviewee some questions about these topics, and the interviewee answers these questions. Generally the interviewer gives no comments on the answers on the spot. Through this interview the interviewer can gain the information about the English proficiency, the normal life, the major, and so on of the interviewee.

In order to simulate the normal graduate job application interview the interviewer's questions are stored according to the topics in a table in the NLDB. For a given topic there are two types of questions. Some can be given randomly, and others should be given in the order. So by dialog simulation the questions are randomly selected or given in sequence. To do this dialog simulation the user can either use the text output, or use speech synthesis by selecting two Microsoft agent character from the fives mentioned above, one representing the user himself (herself), another representing the virtual interviewer. During the simulation all characters lose their chatting patterns defined above.

Some dialogs extracted from a simulated interview between Christoph and a user with the name Petra are shown in the following paragraph.

Petra: good afternoon!
Christoph: Good day! What university do you attend?
Petra: Jilin normal university.
Christoph: When will you graduate from your university?
Petra: 2005.
Christoph: Will you receive any degree?
Petra: yes, i receive english major bachalor degree.
Christoph: What's your major?
Petra: my major is English.
Christoph: Please tell me about the courses you've completed at university.
Petra: I have passed 4 exam for english major student.
Christoph: What is your favorite course?
Petra: my favorate course is English liberary history
Christoph: Which college subjects did you like least? Why?
Petra: maybe the oral English course, because the oral teacher is a foreigner, and his type of teaching is not very good.
Christoph: What is your GRE score?
Petra: sorry,I haven't had this exam
Christoph: Is there anything you regret not having done or would like to have done differently at college?
Petra: pardon
Christoph: Is there anything you regret not having done?
Petra: many things, I think.
Christoph: What are your greatest strengths?
Petra: maybe my listening English.
Christoph: Have you been involved in any extracurricular activities at college?
Petra: yes, I like liberary, so I had been involved in some chinese liberary activity
Christoph: What do you do in your spare time?
Petra: I like play computer game.
Etc.

From the dialog above we can see that in the user input there are some errors or inappropriate expressions. For the goal of real time simulation and out of respect of the interviewee these errors should not be shown during the interview. But the interviewee would like to know whether his response is appropriate and whether he has made mistakes. So after the interview simulation the CSIEC system should analyze the user responses, praise the user for his correct answers, point out the errors and suggest the better answers. By this way the user can learn more practical English expressions used in this kind of discourse. The aim of dialog simulation for specific scenarios can therefore be achieved.

With the function of online syntactic and semantic analysis of the user inputs the CSIEC system, here concretized by the agent Christoph, can display the user the following comments at once after the interview.

Christoph: Ok. We have finished our job application interview. Most of your answers are correct and suitable. Unfortunately there are some spelling or grammar errors, as well as inappropriate expressions. Please read the following sentences carefully, and correct the errors by yourself or with the help of a textbook, or from your teacher and classmates.

*yes, i receive english major bachalor degree.*
*my favorate course is English liberary history*
*yes, I like liberary, so I had been involved in some chinese liberary activity*
*I like play computer game.*

In the current version we'd rather not to replace the role of the teachers in the English education, so the system doesn't point out the real errors, nor correct them. The reason will be interpreted in next parts.

## 3. Features of English Education in China

With the rapid economic, cultural, and educational internationalization in China, and the oncoming of Olympiad 2008 Beijing, the demand for learning a foreign language, first of all, English, is increasing greatly. Therefore English course is now along with Chinese and mathematics treated as one of three core courses in elementary and secondary education, and as one obligated course in higher education.

A student must learn English. Otherwise he or she can not pass any entrance examination to a higher education. For such an obligation course the teacher plays an extremely important role, especially in elementary and secondary education, as the younger students must learn Chinese and English at the same time. Without the instruction of the teachers the students can't learn the fundamental knowledge of English.

The English education in China is traditionally examination oriented, although the experts and authorities have been trying hard to deemphasize the examination score and to emphasize the importance of practical English usage in social communication. According to a report from the Ministration of Education in 2004 there are c.a. 20 millions English learners in China, who must pass an English examination [5].

Chinese, as the most popular mother language in China, differs from English notably in grammar. So the grammar teaching and learning are important in English education. Without basic grammar knowledge the students can't make great progress, as they use English mostly only in school time, and can't learn it unconsciously and spontaneously from the social environment.

The market for English learning is very large, the demand on English education in china is giant, and so many well trained teachers are required. However, on fact there aren't enough English teachers to satisfy this demand, especially those whose mother language is English. The usage of new technologies in English teaching and learning, for example, multimedia computer programs, web-based learning, computer mediated communication, etc., may help solve this problem in some extent. However, an essential pedagogical method can't be realized by computers yet, i.e. the communication in natural language with the learners. This method plays an exceptionally significant role in the language learning, as many language theories and practice have shown [6][7].

The application of CSIEC can meet such requirements, e.g., create a virtual English speaking environment, as we have argued in [1][4]. But its more success relies on its adaptation to the features of English education in China.

## 4. Adapting CSIEC to the Special Chinese Context

One basic function of the CSIEC system, spelling and grammar check, can contribute to improve the grammatical skills of the students. The two new works we have been doing, the selectable chatting partners and the specific dialog simulation, as presented above in this paper, are also our efforts to adapt and localize the CSIEC to English education in China. The former one is hoped to be able to satisfy the various appetites of so many English learners in China and arouse the interests and curiosity of the students. The later is aimed to help the students pass some oral examinations.

Unfortunately few English teachers know this system well. So we are getting in touch with some teachers from elementary, secondary and higher education, and are planning to combine this system with English education in the classroom, and to do a progressive and summative evaluation. With the feedbacks from the students and teachers in hand we believe we can improve the CSIEC system more efficiently and more instruction-oriented. This is the first work we must do next.

Another work is the integration of newest speech recognition technology into the system, so that the user can directly use a microphone to speak with the online chatting partner, and learn to communicate in a more real English learning circumstance. Certainly it will make the syntactic and semantic analysis of the input texts much more difficult.

## 5. References

[1] J. Jia. CSIEC (Computer Simulator in Educational Communication): A Virtual Context-Adaptive Chatting Partner for Foreign Language Learners. In: *Proceedings of ICALT 04*. IEEE Computer Society Press, the USA. 2004. pp.690-692.
[2] http://edu.sina.com.cn/en/2005-03-28/32273.html
[3] http://www.microsoft.com/products/msagent
[4] J. Jia. The study of the application of a web-based chatbot system on the teaching of foreign languages. In: *Proceedings of SITE 04*. AACE Press, the USA. 2004. pp.1201-1207.
[5] Xinhua News Agency. *English education in China*. 2005.
[6] Zhuanglin Hu, etc. *Linguistics: a concise course book*. Peking University Press, China. 2004.
[7] Ruiqing Liu, etc. *Theories and schools of linguistics*. Nanjing Normal University Press, China. 2003.